\documentclass[12pt] {article}
\usepackage{psfrag}
\usepackage{graphicx}
\usepackage{latexsym,amsfonts}
\begin{document}
\setcounter{page}{1}
\def\theequation{\arabic{section}.\arabic{equation}}
\def\theequation{\thesection.\arabic{equation}}
\setcounter{section}{0}

\title{On free massless (pseudo)scalar quantum field theory in
1+1--dimensional space--time}

\author{M. Faber\thanks{E--mail: faber@kph.tuwien.ac.at, Tel.:
+43--1--58801--14261, Fax: +43--1--5864203} ~~and~
A. N. Ivanov\thanks{E--mail: ivanov@kph.tuwien.ac.at, Tel.:
+43--1--58801--14261, Fax: +43--1--5864203}~\thanks{Permanent Address:
State Technical University, Department of Nuclear Physics, 195251
St. Petersburg, Russian Federation}}

\date{\today}

\maketitle
\vspace{-0.5in}
\begin{center}
{\it Atominstitut der \"Osterreichischen Universit\"aten,
Arbeitsbereich Kernphysik und Nukleare Astrophysik, Technische
Universit\"at Wien, \\ Wiedner Hauptstr. 8-10, A-1040 Wien,
\"Osterreich }
\end{center}

\begin{center}
\begin{abstract}
We construct a consistent quantum field theory of a free massless
(pseudo)scalar field in 1+1--dimensional space--times free of infrared
divergences. We show that in such a quantum field theory (i) a
continuous symmetry of (pseudo)scalar field translations is
spontaneously broken, (ii) Goldstone bosons appear as quanta of a free
massless (pseudo)scalar field and (iii) there is a nonvanishing
spontaneous {\it magnetization}. In spite of the existence of a
spontaneous {\it magnetization} the main inequality between vacuum
expectation values of certain operators which have been used for the
derivation of the Mermin--Wagner--Hohenberg theorem (C. Itzykson and
J.--M. Drouffe, {\it STATISTICAL FIELD THEORY}, Vol. I, 1989,
pp.219--224) is fulfilled. 
\end{abstract}
\end{center}

\newpage

\section{Introduction}
\setcounter{equation}{0}

\hspace{0.2in} As has been noted by Klaiber in his seminal paper [1]
devoted to the solution of the massless Thirring model [2,3] within
the operator formalism, the main problem of quantum field theories in
1+1--dimensional space--time is the infrared divergence of the
two--point Wightman functions of a free massless (pseudo)scalar field
$\vartheta(x)$
\begin{eqnarray}\label{label1.1}
\hspace{-0.3in}D^{(+)}(x) &=& \langle
0|\vartheta(x)\vartheta(0)|0\rangle =
\frac{1}{2\pi}\int^{\infty}_{-\infty}\frac{dk^1}{2k^0}\,e^{\textstyle
-\,i\,k\cdot x} = - \frac{1}{4\pi}\,{\ell n}[-\mu^2x^2 +
i\,0\cdot\varepsilon(x^0)], \nonumber\\ \hspace{-0.3in}D^{(-)}(x) &=&
\langle 0|\vartheta(0)\vartheta(x)|0\rangle =
\frac{1}{2\pi}\int^{\infty}_{-\infty}\frac{dk^1}{2k^0}\,e^{\textstyle
+\,i\,k\cdot x} = - \frac{1}{4\pi}\,{\ell n}[-\mu^2x^2 -
i\,0\cdot\varepsilon(x^0)],
\end{eqnarray}
where $\varepsilon(x^0)$ is the sign function, $x^2 = (x^0)^2 -
(x^1)^2$, $k\cdot x = k^0x^0 - k^1x^1$, $k^0 = |k^1|$ is the energy of
a free massless (pseudo)scalar quantum with momentum $k^1$ and $\mu$
is the infrared cut--off reflecting the infrared divergence of the
Wightman function (\ref{label1.1})\,\footnote{Further in order to
underscore the dependence of the two--point Wightman functions of the
scale $\mu$ we will denote $D^{(\pm)}(x) \to D^{(\pm)}(x;
\mu)$.}. This was already stated by Klaiber [1]:``If one wants to
solve the Thirring model, one has to overcome this problem.''

The problem of the infrared divergence of the Wightman function of a
free massless (pseudo)scalar field (\ref{label1.1}) has than been
reformulated by Coleman [4] as the absence of Goldstone bosons and,
correspondingly, of a spontaneously broken continuous symmetry in
1+1--dimensional space--time.

Recently [5] we have shown that Coleman's statement concerning the
absence of a spontaneously broken continuous symmetry in
1+1--dimensional space--time is questionable. Indeed, the main
distinction of a spontaneously broken continuous symmetry from an
unbroken one is in the appearance of massless Goldstone bosons [6].
According to Goldstone's theorem [6], Goldstone bosons accompany the
spontaneous breaking of a continuous symmetry. In this connection
Coleman [4] argued that in a 1+1--dimensional quantum field theory of
a massless (pseudo)scalar field there are no Goldstone bosons.  In
order to prove this statement Coleman considered a quantum field
theory of a free massless (pseudo)scalar field $\vartheta(x)$ with the
Lagrangian
\begin{eqnarray}\label{label1.2}
{\cal L}(x) =
\frac{1}{2}\,\partial_{\mu}\vartheta(x)\partial^{\mu}\vartheta(x),
\end{eqnarray}
where $x = (x^0,x^1)$ is a 2--vector. The equation of motion of the
$\vartheta$--field reads
\begin{eqnarray}\label{label1.3}
\Box \vartheta(x) = 0.
\end{eqnarray}
The Lagrangian (\ref{label1.2}) is invariant under field
translations [5,7] 
\begin{eqnarray}\label{label1.4}
 \vartheta(x) \to \vartheta\,'(x) = \vartheta(x) - 2\,\alpha_{\rm A},
\end{eqnarray}
where $\alpha_{\rm A}$ is an arbitrary parameter. The conserved
current associated with these field translations is equal to
\begin{eqnarray}\label{label1.5}
j_{\mu}(x) = \partial_{\mu}\vartheta(x).
\end{eqnarray}
The total {\it charge} is defined by the time--component of
$j_{\mu}(x)$  [5,7]
\begin{eqnarray}\label{label1.6}
Q(x^0) =\lim_{L\to \infty} \int^{ L/2}_{-
L/2}dx^1\,\frac{\partial}{\partial x^0}\vartheta(x^0,x^1),
\end{eqnarray}
where $L$ is the volume occupied by the system.

It is well--known that the spontaneous breaking of a continuous
symmetry occurs when the ground state of the system is not invariant
under the symmetry group [6,8]. The ground state of the system described
by the Lagrangian (\ref{label1.2}) is not invariant under field
translations (\ref{label1.4}) [5]. Therefore, the field--translation
symmetry should be spontaneously broken and a Goldstone boson should
appear [5].

Therewith, the non--invariance of the ground state of the system can
be demonstrated by acting with the operator $\exp\{-2i\alpha_{\rm
A}Q(0)\}$ on the vacuum wave function $|0\rangle$, i.e. $|\alpha_{\rm
A}\rangle = \exp\{-2i\alpha_{\rm A}Q(0)\}|0\rangle$ [5].

For the calculation of $|\alpha_{\rm A}\rangle$ we use the expansion
of the massless (pseudo)scalar field $\vartheta(x)$ into plane waves
[5]
\begin{eqnarray}\label{label1.7}
\vartheta(x) =
\int^{\infty}_{-\infty}\frac{dk^1}{2\pi}\,\frac{1}{2k^0}\,
\Big(a(k^1)\,e^{\textstyle -i\,k\cdot x} +
a^{\dagger}(k^1)\,e^{\textstyle i\,k\cdot x}\Big),
\end{eqnarray}
where $a(k^1)$ and $a^{\dagger}(k^1)$ are annihilation and creation
operators obeying the standard commutation relation
\begin{eqnarray}\label{label1.8}
[a(k^1), a^{\dagger}(q^1)] = (2\pi)\,2k^0\,\delta(k^1 - q^1).
\end{eqnarray}
From (\ref{label1.6}) we obtain the total {\it charge} operator $Q(0)$
[5]
\begin{eqnarray}\label{label1.9}
Q(0) = -\frac{i}{2}[a(0) - a^{\dagger}(0)].
\end{eqnarray}
Then, we get the wave function $|\alpha_{\rm A}\rangle$
\begin{eqnarray}\label{label1.10}
|\alpha_{\rm A}\rangle = e^{\textstyle -2i\alpha_{\rm A}Q(0)}|0\rangle
  = e^{\textstyle - \alpha_{\rm A}[a(0) - a^{\dagger}(0)]}|0\rangle.
\end{eqnarray}
For the subsequent mathematical operations with the wave functions
$|\alpha_{\rm A}\rangle$ it is convenient to use the regularization
procedure suggested by Itzykson and Zuber [7]. We define the
regularized operator $Q(0)_{\rm R}$ as follows [5]
\begin{eqnarray}\label{label1.11}
Q(0)_{\rm R} = \lim_{L \to
\infty}\int^{\infty}_{-\infty}dx^1\,\frac{\partial}{\partial
x^0}\vartheta(x^0,x^1)\Bigg|_{x^0=0}e^{\textstyle - (x^1)^2/L^2}.
\end{eqnarray}
The regularized wave function $|\alpha_{\rm A}\rangle_{\rm R}$ reads
then
\begin{eqnarray}\label{label1.12}
|\alpha_{\rm A}\rangle_{\rm R} &=& e^{\textstyle -2i\alpha_{\rm
  A}Q(0)_{\rm R}}|0\rangle = \nonumber\\ &=&\lim_{L \to
  \infty}\exp\Big\{ -\frac{\alpha_{\rm
  A}L}{2\sqrt{\pi}}\int^{\infty}_{-\infty}dk^1[a(k^1) -
  a^{\dagger}(k^1)]\,e^{\textstyle - L^2(k^1)^2/4}\Big\}|0\rangle.
\end{eqnarray}
The normal--ordered energy operator of the massless (pseudo)scalar
field described by the Lagrangian (\ref{label1.2}) is equal to
\begin{eqnarray}\label{label1.13}
\hspace{-0.3in}&&\hat{H}(x^0) = \int^{\infty}_{-\infty}dx^1\,{\cal
H}(x^0,x^1)
=\int^{\infty}_{-\infty}dx^1\,[:\pi(x^0,x^1)\dot{\vartheta}(x^0,x^1)
- {\cal L}(x^0,x^1):] = \nonumber\\ \hspace{-0.3in}&&=
\frac{1}{2}\int^{\infty}_{-\infty}dx\,:\Bigg[\Bigg(\frac{\partial
\vartheta(x^0,x^1)}{\partial x^0}\Bigg)^2 + \Bigg(\frac{\partial
\vartheta(x^0,x^1)}{\partial x^1}\Bigg)^2\Bigg]: =
\frac{1}{2}\int^{\infty}_{-\infty}\frac{dk^1}{2\pi}\,a^{\dagger}(k^1)a(k^1),
\end{eqnarray}
where $\pi(x) = \dot{\vartheta}(x)$ is the conjugate momentum and
$\dot{\vartheta}(x)$ is the time derivative.

One can easily show that the wave functions $|\alpha_{\rm
A}\rangle_{\rm R}$ are eigenfunctions of the energy operator
(\ref{label1.13}) for the eigenvalue zero
\begin{eqnarray}\label{label1.14}
\hat{H}(x^0)|\alpha_{\rm A}\rangle_{\rm R} = E(\alpha_{\rm
A})|\alpha_{\rm A}\rangle_{\rm R} = 0.
\end{eqnarray}
This shows that the energy of the vacuum state is infinitely
degenerate, and the vacuum state depends on the field translations
(\ref{label1.4}). The wave functions of the vacuum state $|\alpha_{\rm
A}\rangle_{\rm R}$ and $|\alpha'_{\rm A}\rangle_{\rm R}$ are not
orthogonal to each other for $\alpha'_{\rm A}\not= \alpha_{\rm A}$ and
the scalar product ${_{\rm R}\langle}\alpha'_{\rm A}|\alpha_{\rm
A}\rangle_{\rm R}$ amounts to [5]
\begin{eqnarray}\label{label1.15}
{_{\rm R}\langle}\alpha'_{\rm A}|\alpha_{\rm A}\rangle_{\rm R} =
e^{\textstyle - (\alpha'_{\rm A} - \alpha_{\rm A})^2}.
\end{eqnarray}
Since the eigenvalue of the wave functions $|\alpha_{\rm A}\rangle$ is
zero, they can be orthogonalized by any appropriate orthogonalization
procedure as used in molecular and nuclear physics.

We would like to emphasize that the results expounded above are not
related to the impossibility to determine the two--point Wightman
function (\ref{label1.1}) in the infrared region of $\vartheta$--field
fluctuations.  In fact, the analysis of the non--invariance of the
vacuum wave function under the symmetry transformations
(\ref{label1.4}) treats the massless (pseudo)scalar field at the tree
level. This is an appropriate description, since the massless
(pseudo)scalar field $\vartheta(x)$ is free, no vacuum fluctuations are
entangled and the quanta of the massless $\vartheta$--field are
on--mass shell.

Following the Itzykson--Zuber analysis of the 1+1--dimensional
massless (pseudo)scalar field theory of the $\vartheta$--field
described by the Lagrangian (\ref{label1.2}) one can show [5] that the
translation symmetry (\ref{label1.4}) is spontaneously broken: (i) the
ground state is not invariant under the field--translation symmetry,
(ii) the energy of the ground state is infinitely degenerated and
(iii) Goldstone bosons appear and they are the quanta of the
$\vartheta$--field. Hence, all requirements for a continuous symmetry
to be spontaneously broken are available in the 1+1--dimensional
quantum field theory of a massless (pseudo)scalar field described by
the Lagrangian (\ref{label1.2}).

The paper is organized as follows. In section 2 we show that the
generating functional of Green functions of a free massless
(pseudo)scalar field $\vartheta(x)$ does not depend on the infrared
cut--off and the two--point causal and Wightman functions can be made
finite in the infrared region. In section 3 we consider the
low--frequency quanta of a free massless (pseudo)scalar field as an
ensemble which is described by a wave function of a coherent state
with a random fluctuating field $\eta(k^1)$ [9]. We suggest to treat
this random field as a hidden parameter of the theory. We assume that
all quantities such as correlation functions defined in the quantum
field theory of a free massless (pseudo)scalar field $\vartheta(x)$
should be averaged over this parameter. In this way we derive Wightman
functions which are non--singular in the infrared region. In the
Conclusion we discuss the relation of the infrared regularized quantum
field theory of a free massless (pseudo)scalar field $\vartheta(x)$ to
the Mermin--Wagner--Hohenberg (MWH) theorem [10] stating the vanishing
of the {\it long--range order} and a spontaneous {\it magnetization}
in two dimensional field theories. We show that in the quantum field
theory of the free massless (pseudo)scalar field free of infrared
divergences there is a non--vanishing spontaneous {\it magnetization}.
We demonstrate that in spite of the non--vanishing value of the
spontaneous {\it magnetization} the main inequality which has been
used for the derivation of the MWH theorem is fulfilled. We argue that
this result cannot be a counterexample to the MWH theorem, since this
theorem was formulated only for non--zero temparature $T \neq 0$,
whereas a spontaneous {\it magnetization} in the quantum field theory
of the free massless (pseudo)scalar field is calculated for $T=0$. We
accentuate that the infrared regularized quantum field theory of a
free massless (pseudo)scalar field $\vartheta(x)$ agrees well with the
results obtained in Ref.[5] for the solution of the massless Thirring
model with fermion fields quantized in the chirally broken phase.

\section{Generating functional of Green functions of a free
massless (pseudo)scalar field} 
\setcounter{equation}{0}

\hspace{0.2in} It is well--known that the solution of a quantum field
theory corresponds to the evaluation of any correlation function. In
the quantum field theory of a free massless (pseudo)scalar field
$\vartheta(x)$ any correlation function can be evaluated by means of
the generating functional of Green functions defined by
\begin{eqnarray}\label{label2.1}
\hspace{-0.15in}Z[J] = \Big\langle 0\Big|{\rm T}\Big(e^{\textstyle
i\int d^2x\,\vartheta(x)J(x)}\Big)\Big|0\Big\rangle = \int {\cal
D}\vartheta\,e^{\textstyle i\int
d^2x\,\Big[\frac{1}{2}\,\partial_{\mu}\vartheta(x)\partial^{\mu}
\vartheta(x) + \vartheta(x)J(x)\Big]},
\end{eqnarray}
where ${\rm T}$ is a time--ordering operator and $J(x)$ is an external
source of the free massless (pseudo)scalar field $\vartheta(x)$.

In terms of $Z[J]$ an arbitrary correlation function of the
$\vartheta$--field can be defined as follows
\begin{eqnarray}\label{label2.2}
&&G(x_1,\ldots,x_n;y_1,\ldots,y_p) = \langle
0|F(\vartheta(x_1),\ldots,\vartheta(x_n);\vartheta(y_1),\ldots,\vartheta(y_p))|0\rangle
= \nonumber\\ &&= F\Big(-i\frac{\delta}{\delta
J(x_1)},\ldots,-i\frac{\delta}{\delta J(x_n)};- i\frac{\delta}{\delta
J(y_1)},\ldots,-i\frac{\delta}{\delta
J(y_p)}\Big)Z[J]\Big|_{\textstyle J = 0}.
\end{eqnarray}
As usual in the sine--Gordon, Thirring and $XY$ models [11--16] one
encounters the problem of the evaluation of correlation functions of
the following kind
\begin{eqnarray}\label{label2.3}
&&G(x_1,\ldots,x_n;y_1,\ldots,y_p) = \Big\langle 0\Big|{\rm
T}\Big(\prod^n_{j=1}e^{\textstyle +
i\beta\vartheta(x_j)}\prod^p_{k=1}e^{\textstyle
-i\beta\vartheta(y_k)}\Big)\Big|0\Big\rangle=\nonumber\\
&&=\exp\Big\{-i\beta\sum^n_{j = 1}\frac{\delta}{\delta J(x_j)} +
i\beta\sum^p_{k = 1}\frac{\delta}{\delta
J(y_k)}\Big\}Z[J]\Big|_{\textstyle J = 0}.
\end{eqnarray}
Since the path--integral over the $\vartheta$--field (\ref{label2.1})
is Gaussian, it can be evaluated explicitly. The result reads
\begin{eqnarray}\label{label2.4}
Z[J] = \lim_{\textstyle \mu \to 0}\exp\,\Big\{i\,\frac{1}{2}\int
d^2x\,d^2y\,J(x)\,\Delta(x-y;\mu)\,J(y)\Big\},
\end{eqnarray}
where $\Delta(x-y;\mu)$, the causal two--point Green function, obeys
the equation
\begin{eqnarray}\label{label2.5}
\Box\,\Delta(x-y;\mu) = \delta^{(2)}(x-y)
\end{eqnarray}
and is given by the expression 
\begin{eqnarray}\label{label2.6}
\Delta(x-y;\mu) =-\,\frac{i}{4\pi}\,{\ell n}[-\mu^2(x-y)^2 + i\,0\,],
\end{eqnarray}
where $\mu$ is the infrared cut--off that should be taken finally in
the limit $\mu \to 0$. 

The presence of the infrared cut--off is related to the infrared
problem of a free massless (pseudo)scalar field theory formulated by
Klaiber [1] and Coleman [4]. The removal of the infrared cut--off from
the Green function $\Delta(x-y;\mu)$ and its replacement by a finite
scale $M$ should solve this infrared problem.

In order to understand the behaviour of $Z[J]$ in the limit $\mu \to
0$ we suggest to factorize the contribution of the infrared cut--off
$\mu$ by introducing a finite arbitrary scale $M$. This yields
\begin{eqnarray}\label{label2.7}
Z[J] &=& \lim_{\textstyle \mu \to 0}\,\exp\,\Big\{\frac{1}{8\pi}\int
d^2x\,d^2y\,J(x)\,{\ell n}[-\mu^2(x-y)^2 + i\,0\,]\,J(y)\Big\} =
\nonumber\\ &=&\exp\,\Big\{\frac{1}{8\pi}\int d^2x\,d^2y\,J(x)\,{\ell
n}[-M^2(x-y)^2 + i\,0\,]\,J(y)\Big\}\nonumber\\
&&\times\lim_{\textstyle \mu \to 0}\,\exp\Big\{-\,\frac{1}{8\pi}{\ell
n}\frac{M^2}{\mu^2}\Big(\int d^2x\,J(x)\Big)^2\Big\}=\lim_{\textstyle
\mu \to 0}\Bigg(\frac{\mu}{M}\Bigg)^{\textstyle \frac{\textstyle
1}{\textstyle 4\pi}\Big(\int d^2x\,J(x)\Big)^2} \nonumber\\ &&\times
\,\exp\,\Big\{\frac{1}{8\pi}\int d^2x\,d^2y\,J(x)\,{\ell
n}[-M^2(x-y)^2 + i\,0\,]\,J(y)\Big\}.
\end{eqnarray}
Since the power of the ratio $\mu/M$ is always positive, so for any
arbitrary external source $J(x)$ with a non--vanishing
1+1--dimensional volume integral
\begin{eqnarray}\label{label2.8}
\int d^2x\, J(x) \neq 0
\end{eqnarray}
the generating functional $Z[J]$ vanishes in the limit $\mu \to 0$,
i.e. $Z[J] = 0$. 

In the Schwinger formulation of quantum field theory [17] the
generating functional of Green functions $Z[J]$ given by
(\ref{label2.4}) defines the amplitude of the {\it vacuum--vacuum}
transition $\langle 0_+|0_-\rangle^J$, i.e. $Z[J] = \langle
0_+|0_-\rangle^J$. The vanishing of $Z[J]$ corresponds to the
vanishing of the amplitude of the {\it vacuum--vacuum} transition
$\langle 0_+|0_-\rangle^J = 0$. According to Schwinger [17] a quantum
field theory with $\langle 0_+|0_-\rangle^J = 0$ has no meaning.

In order to make a quantum field theory of a free massless
(pseudo)scalar field $\vartheta(x)$ meaningful we have to get a
non--vanishing generating functional of Green functions $Z[J]$. This
can be obtained by imposing the constraint
\begin{eqnarray}\label{label2.9}
\int d^2x\,J(x) = 0.
\end{eqnarray}
We would like to emphasize that due to this constraint $Z[J]$ becomes
invariant under the $\vartheta$--field translations (\ref{label1.4})
as well as the Lagrangian (\ref{label1.2}).

For the massless pseudoscalar field $\vartheta(x)$ the constraint
(\ref{label2.9}) is fulfilled automatically. Indeed, due to the
conservation of parity the external source of the field $\vartheta(x)$
should obey the relation $J(x^0,x^1) = - J(x^0,-x^1)$. For a scalar
$\vartheta$--field, when $J(x^0,x^1) = J(x^0,-x^1)$, the relation
(\ref{label2.2}) can be fulfilled for a rather broad class of
analytical and generalized functions [18].

Recall that the constraint (\ref{label2.9}) has been implicitly used
for the evaluation of correlation functions (\ref{label2.3}) which can
be transcribed as follows
\begin{eqnarray}\label{label2.10}
\hspace{-0.3in}G(x_1,\ldots,x_n;y_1,\ldots,y_p) &=& \Big\langle
0\Big|{\rm T}\Big(e^{\textstyle i\int
d^2x\,\vartheta(x)J(x;x_1,\ldots,x_n;y_1\ldots,y_p)}\Big)\Big|0\Big\rangle,
\end{eqnarray}
where the current $J(x)$ is defined by 
\begin{eqnarray}\label{label2.11}
J(x;x_1,\ldots,x_n;y_1\ldots,y_p) = \beta\sum^n_{j=1}\delta^{(2)}(x -
x_j) - \beta\sum^p_{k=1}\delta^{(2)}(x - y_k).
\end{eqnarray}
Substituting (\ref{label2.11}) in (\ref{label2.9}) we obtain
\begin{eqnarray}\label{label2.12}
\int d^2x\,J(x;x_1,\ldots,x_n;y_1\ldots,y_p) = \beta (n-p).
\end{eqnarray}
It is well--known [11--16] that for $n = p$ the correlation functions
(\ref{label2.10}) do not depend on the infrared cut--off, in turn, for
$n\neq p$ the dependence of the correlation functions on $\mu$ leads
to their vanishing in the limit $\mu \to 0$. Unfortunately, nobody
made an attempt to relate this result with the vanishing of $Z[J]$ in
the limit $\mu \to 0$ for an arbitrary defined external source $J(x)$.

As has been shown in [16] the constraint (\ref{label2.9}) has turned
out to be rather useful for the proof of non--perturbative
renormalizability of the sine--Gordon model described by the
Lagrangian [5,11,16]
\begin{eqnarray}\label{label2.13}
{\cal L}(x) =
\frac{1}{2}\,\partial_{\mu}\vartheta(x)\partial^{\mu}\vartheta(x) +
\frac{\alpha}{\beta^2}\,(\cos\beta\vartheta(x) - 1),
\end{eqnarray}
where $\alpha$ and $\beta$ are parameters of the model [5,11,16]. The
main peculiarity of the sine--Gordon model is the existence of soliton
solutions having the properties of fermions [19,20] (see also [5]).

Hence, in order to make the quantum field theory of a free massless
(pseudo)scalar field meaningful the constraint (\ref{label2.9}) should
be imposed. According to this constraint the generating functional of
Green functions $Z[J]$ reads
\begin{eqnarray}\label{label2.14}
Z[J] = \exp\,\Big\{\frac{1}{8\pi}\int d^2x\,d^2y\,J(x)\,{\ell
n}[-M^2(x-y)^2 + i\,0\,]\,J(y)\Big\}.
\end{eqnarray}
In this form the generating functional of Green functions $Z[J]$ is
well--defined. Thus, the constraint (\ref{label2.2}) provides the
independence of the generating functional of Green functions of the
free massless (pseudo)scalar field $\vartheta(x)$ of the infrared
cut--off. This makes the quantum field theory described by $Z[J]$ in
(\ref{label2.14}) meaningful.

Of course, a certain cautiousness is needed because constraints on the
external sources $J(x)$ may lead to a loss of information about the
response of the quantum system. Therefore, one has to be convinced
that in the case of the free massless (pseudo)scalar quantum field
theory, defined in 1+1--dimensional space--time, there is no a loss of
information about the response of the free massless (pseudo)scalar
field to the external perturbations restricted by the constraint
(\ref{label2.9}).

The physical meaning of the constraint (\ref{label2.9}) can be easily
understood in the momentum representation, where the (pseudo)scalar
field $\vartheta(x)$ and the external source $J(x)$ are expressed by
their Fourier transforms $\tilde{\vartheta}(k)$ and $\tilde{J}(k)$
\begin{eqnarray}\label{label2.15}
\vartheta(x) = \int
\frac{d^2k}{(2\pi)^2}\,\tilde{\vartheta}(k)\,e^{\textstyle -ik\cdot
x}\quad,\quad J(x) = \int
\frac{d^2k}{(2\pi)^2}\,\tilde{J}(k)\,e^{\textstyle -ik\cdot x}.
\end{eqnarray}
By making a change of variables $\vartheta(x) \to
\tilde{\vartheta}(k)$ the generating functional of Green function
acquires the form
\begin{eqnarray}\label{label2.16}
Z[J] = \int {\cal D}\tilde{\vartheta}\,
\exp\Big\{i\int\frac{d^2k}{(2\pi)^2}\,\Big[\frac{1}{2}\,(k^2+i\,0)
\tilde{\vartheta}(k) \tilde{\vartheta}(-k) +
\tilde{\vartheta}(k)\tilde{J}(-k)\Big]\Big\}.
\end{eqnarray}
Integrating over the $\tilde{\vartheta}$--field we arrive at the
expression
\begin{eqnarray}\label{label2.17}
Z[J] &=& \exp\,\Big\{- i\,\frac{1}{2}\int
\frac{d^2k}{(2\pi)^2}\,\frac{\tilde{J}(k)\tilde{J}(-k)}{k^2 +
i\,0}\,\Big\} = \nonumber\\ &=& \exp\,\Big\{- i\,\frac{1}{2}\int
d^2x\,d^2y\,J(x)\int \frac{d^2k}{(2\pi)^2}\,\frac{e^{\textstyle
ik(x-y)}}{k^2 + i\,0}\,J(y)\Big\}.
\end{eqnarray}
Integrating over $k^0$ we reduce the exponent to the form
\begin{eqnarray}\label{label2.18}
\hspace{-0.5in}&&Z[J] =\exp\,\Big\{-
\frac{1}{4\pi}\int^{\infty}_{-\infty}
\frac{dk^1}{2|k^1|}\,\tilde{J}(-|k^1|,k^1)\tilde{J}(|k^1|,-k^1)\Big\}=
\nonumber\\
\hspace{-0.5in}&&=\exp\,\Big\{- \frac{1}{4\pi}\int
d^2x\,d^2y\,J(x)\,\Big[\theta(x^0 - y^0)\int^{\infty}_{-\infty}
\frac{dk^1}{2|k^1|}\,e^{\textstyle -i|k^1|(x^0 - y^0) + ik^1(x^1 -
y^1)}\nonumber\\
\hspace{-0.5in}&& + \theta(y^0 - x^0)\int^{\infty}_{-\infty}
\frac{dk^1}{2|k^1|}\,e^{\textstyle +i|k^1|(x^0 - y^0) - ik^1(x^1 -
y^1)}\Big]\,J(y)\Big\}.
\end{eqnarray}
Substituting the external source $J(x)$ defined by the momentum
integral into (\ref{label2.15}) we obtain $\tilde{J}(0) = 0$. Hence,
the constraint (\ref{label2.9}) is equivalent to the vanishing of the
Fourier transform of the external source for zero 2--momentum $k = 0$.
Due to this the momentum integrals become convergent in the infrared
region $k \to 0$. As result the generating functional $Z[J]$ does not
depend on the infrared cut--off $\mu$. This makes $Z[J]$
well--defined.

It is obvious that in $n\ge 3$ dimensional space--time, where the
generating functional of Green functions $Z[J]$ is defined by
\begin{eqnarray}\label{label2.19}
Z[J]= \exp\,\Big\{- i\,\frac{1}{2}\int
\frac{d^nk}{(2\pi)^2}\,\frac{\tilde{J}(k)\tilde{J}(-k)}{k^2 +
i\,0}\,\Big\},
\end{eqnarray}
the problem of the ill--definition of the momentum integral in the
infrared region does not appear due to the contribution of the measure
$d^nk \propto k^{n-1}dk$.

In order to clarify the physical meaning of the constraint
$\tilde{J}(0) = 0$ let us consider the product
$\tilde{\vartheta}(k)\tilde{J}(-k)$ in (\ref{label2.16}). Due to
$\tilde{J}(0) = 0$ we get $\tilde{\vartheta}(0)\tilde{J}(0) = 0$. This
means that the zero--mode configuration $\tilde{\vartheta}(0)$ of the
$\vartheta$--field does not couple to an external source. In other
word setting $\tilde{J}(0) = 0$ we do not excite the zero--mode
configuration of the $\vartheta$--field. As a result this
configuration does not contribute to $Z[J]$ or differently to the
amplitude of the {\it vacuum--vacuum} transition $\langle
0_+|0_-\rangle^J$.

Such a treatment of the contribution of the zero--mode configuration
of the free massless (pseudo)scalar field is equivalent to some extent
to the procedure suggested by Hasenfratz [21]\,\footnote{We are
grateful to Oleg Borisenko for calling our attention to Hasenfratz's
paper [21].}  to treat the zero--mode configuration of self--coupled
massless scalar fields in lattice $\sigma$--models with $O(N)$
symmetry defined in one and two dimensions in finite and infinite
volumes.

The role of the zero--mode configuration of the $\vartheta$--field can
be understood by analysing a mechanical analogy of a massless
(pseudo)scalar field. According to [22] the continuous system
described by the Lagrangian (\ref{label1.2}) is equivalent to a
one--dimensional chain of $N$ harmonic oscillators with equal masses,
equal equilibrium separations and a potential energy taking into
account only nearest neighbors. Their motion can be described by
displacements $q_i\,(i=1,\ldots, N)$ which couple to the external
sources as $\sum^N_{i=1}q_iJ_i$. In the representation of normal
coordinates this system reduces to the set of $N-1$ decoupled normal
non--zero frequency mode configurations and one zero--mode
configuration proportional to the displacement of the center of mass
of $N$ coupled harmonic oscillators $Q_0 \propto \sum^N_{i=1}q_i$. The
coupling of this zero--mode configuration to the external sources is
proportional to $Q_0\sum^N_{i=1}J_i$. Since at $J_i = 0\,(i =
1,\ldots,N)$ the zero--mode configuration does not affect the
evolution of the system, one can exclude a motion of the center of
mass even for $J_i \neq 0\,(i = 1,\ldots,N)$ by the constraint
$\sum^N_{i=1}J_i = 0$ which corresponds to (\ref{label2.9}).  

This clarifies the physical meaning of the constraint
(\ref{label2.9}), that is, the removal of the zero--mode configuration
related to the motion of the center of mass of $N$ coupled harmonic
oscillators, which in the continuous limit and at $N \to \infty$ are
described by the Lagrangian (\ref{label1.2}).  For a free quantum
system the exclusion of a collective motion of a system as a
structureless configuration does not affect the evolution of the
system caused by a relative motion in it and, of course, does not lead
to a loss of an important information about the response to
perturbations induced by external sources.

For a system of coupled quantum fields such an exclusion would not be
innocent. However, as has been shown by Hasenfratz [21] that the
exclusion of the zero--mode configuration of massless self--coupled
scalar fields in one and two dimensions in finite and infinite volumes
described by the $\sigma$--models with $O(N)$ internal symmetry allows
to construct a self--consistent and well--defined perturbation theory.

The inclusion of a finite scale $M$ instead of the infinitesimal scale
$\mu$ leads to Fourier transform of the Wightman function free of
infrared divergences. In order to confirm this statement we suggest to
transcribe the r.h.s. of (\ref{label2.14}) in the form
\begin{eqnarray}\label{label2.20}
Z[J] = \exp\,\Big\{\frac{i}{2}\int
d^2x\,d^2y\,J(x)\,\Delta(x-y; M)\,J(y)\Big\},
\end{eqnarray}
where $\Delta(x-y; M)$ is given by (\ref{label2.6}) with the
replacement $\mu \to M$ and obeys the equation (\ref{label2.5}).

Now let us show that due to the finite scale $M$ the Wightman
functions (\ref{label1.1}) become convergent in the infrared region,
$k^1 \to 0$. As the causal Green function $\Delta(x;M)$ is related to
the Wightman functions $D^{(\pm)}(x; M)$ by the standard relation [14]
\begin{eqnarray}\label{label2.21}
\Delta(x;M) = i\,\theta(+ x^0)\,D^{(+)}(x; M) +
i\,\theta(-x^0)\,D^{(-)}(x; M),
\end{eqnarray}
the Wightman functions $D^{(\pm)}(x; M)$ are equal to
\begin{eqnarray}\label{label2.22}
D^{(\pm)}(x;M) =- \frac{1}{4\pi}\,{\ell n}[-M^2x^2 \pm
i\,0\cdot\varepsilon(x^0)].
\end{eqnarray}
Since the r.h.s. of (\ref{label2.22}) can be treated as a limit
\begin{eqnarray}\label{label2.23}
D^{(\pm)}(x;M) =\frac{1}{2\pi}\lim_{\textstyle \mu \to
0}\Big(K_0(\mu \sqrt{- x^2\pm i\,0\cdot\varepsilon(x^0)}) -
K_0(\mu\,\lambda_{\rm M})\Big),
\end{eqnarray}
where we have denoted $\lambda_{\rm M} = 1/M$, the Fourier transforms
of Wightman functions are defined by (see Appendix C of Ref.[16])
\begin{eqnarray}\label{label2.24}
\hspace{-0.3in}&&D^{(\pm)}(x;M) =\lim_{\textstyle \mu\to
0}\frac{1}{2\pi}\int^{\infty}_{-\infty}\frac{dk^1}{2\sqrt{(k^1)^2 +
\mu^2}}\Big(e^{\textstyle \mp \,i\sqrt{(k^1)^2 + \mu^2}\,x^0 \pm
\,i\,k^1x^1} - \cos(k^1\lambda_{\rm M})\Big).\nonumber\\
\hspace{-0.3in}&&
\end{eqnarray}
Here we have used the integral representations of the McDonald
function $K_0(z)$ [23].

Taking the limit $\mu \to 0$ we obtain
\begin{eqnarray}\label{label2.25}
\hspace{-0.3in}D^{(\pm)}(x;M)
=\frac{1}{2\pi}\int^{\infty}_{-\infty}\frac{dk^1}{2k^0}
\Big(e^{\textstyle \mp \,i\,k\cdot x} - \cos(k^1\lambda_{\rm
M})\Big).
\end{eqnarray}
Thus, the Wightman functions (\ref{label2.25}) are obviously
convergent in the infrared region $k^1 \to 0$. This solves the {\it
infrared} problem of the free massless (pseudo)scalar field theory
pointed out by Klaiber [1] and Coleman [4].

\section{Coherent states of low--frequency quanta of a free
massless (pseudo)scalar field} 
\setcounter{equation}{0}

\hspace{0.2in} The knowledge of the generating functional of Green
functions $Z[J]$ is sufficient for the evaluation of any correlation
function in the quantum field theory under consideration. According to
(\ref{label2.2}) and (\ref{label2.3}) the evaluation of a correlation
function reduces to the calculation of functional derivatives of
$Z[J]$ with respect to the external source $J(x)$. One can easily show
that the use of the constraint (\ref{label2.9}) during the variation
with respect to the external source $J(x)$ does not affect the final
result. Therefore, the analysis of Wightman functions in terms of
vacuum expectation values of the (pseudo)scalar field $\vartheta(x)$
is no more required.

Nevertheless, to make our consideration of the infrared divergences in
the free massless (pseudo)scalar field theory more complete we suggest
a variant of the quantum field theory of the free massless
(pseudo)scalar field reproducing the two--point Wightman functions
(\ref{label2.25}) by means of the redefinition of the (pseudo)scalar
field $\vartheta(x)$.

For the derivation of the term $\cos(k^1\lambda_{\rm M})$ in
(\ref{label2.25}) at the level of the redefinition of a quantum field
$\vartheta(x)$ we would like to notice that the wave function
$|\alpha_{\rm A}\rangle_{\rm R}$ (\ref{label1.12}) of the vacuum state
looks like a wave function of a canonical coherent state produced by a
unitary transformation from a fiducial vacuum state $|0\rangle$
[9]. Following this similarity we suggest to describe collective
states of low--frequency quanta of the free massless (pseudo)scalar
field $\vartheta(x)$, defined by the Lagrangian (\ref{label1.2}), in
terms of coherent states [9].

A collective state of low--frequency quanta of the massless scalar field
$\vartheta(x)$ we suggest to represent by a wave function
\begin{eqnarray}\label{label3.1}
|\eta\rangle = e^{\textstyle \hat{Q}[\eta]}|0\rangle.
\end{eqnarray}
The operator $\hat{Q}[\eta]$ is defined by
\begin{eqnarray}\label{label3.2}
\hat{Q}[\eta]= \int^{\infty}_{-\infty}\frac{dk^1}{2\pi}\,
 \frac{\eta(k^1)}{2k^0}\,[a(k^1) - a^{\dagger}(k^1)],
\end{eqnarray}
where $\eta(k^1)$ is an arbitrary function which we treat as a random
field variable. The $\eta(k^1)$--field is concentrated in the infrared
region close to $k^1 \to 0$ and is almost zero everywhere for finite
momenta. For example, in the case of the vacuum wave functions
$|\alpha_{\rm A}\rangle_{\rm R}$ the $\eta(k^1)$--field is defined by
Gaussian--like functions
\begin{eqnarray}\label{label3.3}
\eta_{\textstyle \,_{\textstyle \alpha_{\rm A}}}(k^1) =
-2\sqrt{\pi}\alpha_{\rm A}L\,|k^1|\, e^{\textstyle -L^2(k^1)^2/4},
\end{eqnarray}
where the spatial volume of the system $L$ should tend to infinity,
$L\to \infty$.

The wave function $|\eta\rangle$ is normalized to unity
\begin{eqnarray}\label{label3.4}
\langle \eta|\eta \rangle = 1.
\end{eqnarray}
The wave function (\ref{label3.1}) is constructed in complete analogy
to the vacuum wave functions $|\alpha_{\rm A}\rangle_{\rm R}$
related to field translations (\ref{label1.4}) caused by chiral
rotations of massless Thirring fermions [5].

The operators of annihilation and creation obey the relations
\begin{eqnarray}\label{label3.5}
\hspace{-0.3in}&& e^{\textstyle \hat{Q}[\eta]}a(k^1) e^{\textstyle
-\hat{Q}[\eta]} = a(k^1) + \eta(k^1),\nonumber\\ \hspace{-0.3in}&&
e^{\textstyle \hat{Q}[\eta]}a^{\dagger}(k^1) e^{\textstyle
-\hat{Q}[\eta]} = a^{\dagger}(k^1) + \eta(k^1).
\end{eqnarray}
This implies that the wave function $|\eta\rangle$ is the
eigenfunction of the annihilation operator $a(k^1)$ with the
eigenvalue $\eta(k^1)$ [9]
\begin{eqnarray}\label{label3.6}
a(k^1)|\eta\rangle = \eta(k^1)|\eta\rangle.
\end{eqnarray}
Using the relation
\begin{eqnarray}\label{label3.7}
e^{\textstyle A + B} = e^{\textstyle -\frac{1}{2}\,[A,B]}\,e^{\textstyle
A}\,e^{\textstyle B}
\end{eqnarray}
one can reduce the wave function $|\eta\rangle$ of (\ref{label3.1}) to
the standard form of a coherent state [9]
\begin{eqnarray}\label{label3.8}
|\eta\rangle =
 \exp\Big\{-\frac{1}{2\pi}\int^{\infty}_{-\infty}\frac{dk^1}{2k^0}\,
 \frac{1}{2}\,\eta^2(k^1)\Big\}
 \exp\Big\{\int^{\infty}_{-\infty}\frac{dk^1}{2\pi}\,
 \frac{\eta(k^1)}{2k^0}\,a^{\dagger}(k^1)\Big\}|0\rangle.
\end{eqnarray}
The completeness condition can be represented by a path integral over
the $\eta$--field [9]
\begin{eqnarray}\label{label3.9}
\hat{1} = \int {\cal D}\eta\,|\eta\rangle\langle \eta|,
\end{eqnarray}
where $\hat{1}$ is a unit operator. Using the completeness condition
and the normalization of the fiducial vacuum state wave function
$\langle 0|0\rangle = 1$ we obtain
\begin{eqnarray}\label{label3.10}
\langle 0|0\rangle = \int {\cal D}\eta\,\langle 0|\eta\rangle \langle
\eta|0\rangle = \int {\cal D}\eta
\exp\Big\{-\frac{1}{2\pi}\int^{\infty}_{-\infty} \frac{dk^1}{2k^0}\,
\eta^2(k^1)\Big\} = 1.
\end{eqnarray}
The exponent of the $c$--number factor in (\ref{label3.10}) is related
to the number of quanta of the massless (pseudo)scalar
$\vartheta$--field in the $|\eta \rangle$ state. In terms of creation
and annihilation operators the operator $\hat{N}$ of the number of
quanta reads
\begin{eqnarray}\label{label3.11}
\hat{N} = \frac{1}{2\pi}\int^{\infty}_{-\infty}\frac{dk^1}{2k^0}\,
a^{\dagger}(k^1)a(k^1).
\end{eqnarray}
Using (\ref{label3.6}) we obtain
\begin{eqnarray}\label{label3.12}
N[\eta] = \langle \eta|\hat{N}|\eta \rangle = \frac{1}{2\pi}
\int^{\infty}_{-\infty}\frac{dk^1}{2k^0}\,\eta^2(k^1).
\end{eqnarray}
The energy of the state $|\eta \rangle$ is equal to 
\begin{eqnarray}\label{label3.13}
E[\eta] = \langle \eta|\hat{H}|\eta \rangle =\frac{1}{2}
\int^{\infty}_{-\infty}\frac{dk^1}{2\pi}\,\eta^2(k^1).
\end{eqnarray}
The action of the creation operator $a^{\dagger}(k^1)$ on the state
$|\eta\rangle$ can be obtained by using (\ref{label3.8}) and reads
\begin{eqnarray}\label{label3.14}
\hspace{-0.3in}a^{\dagger}(k^1)|\eta\rangle =
\exp\Big\{-\frac{1}{2\pi}\int^{\infty}_{-\infty}\frac{dq^1}{2q^0}
\frac{1}{2}\eta^2(q^1)\Big\}
\exp\Big\{\frac{1}{2\pi}\int^{\infty}_{-\infty}\frac{dq^1}{2q^0}
\eta(q^1)a^{\dagger}(q^1)\Big\}a^{\dagger}(k^1)|0\rangle.
\end{eqnarray}
The r.h.s. of this relation can be rewritten in the form of a
variational derivative with respect to $\eta(k^1)$
\begin{eqnarray}\label{label3.15}
a^{\dagger}(k^1)|\eta\rangle = \Big(\eta(k^1) + \frac{\delta}{\delta
\eta(k^1)}\Big)|\eta\rangle,
\end{eqnarray}
where we have used the definition
\begin{eqnarray}\label{label3.16}
\frac{\delta \eta(q^1)}{\delta \eta(k^1)} = (2\pi)\,2k^0\,\delta(k^1 -
q^1).
\end{eqnarray}
Since $\eta(k^1)$ is an auxiliary intrinsic, some kind of hidden,
parameter of a free massless (pseudo)scalar field theory, all
correlation functions should be averaged over the $\eta$--field
fluctuations. Such an average we represent in the form of the
path--integral over the $\eta$--field fluctuations normalized by the
condition (\ref{label3.10}).

Using the random $\eta$--field we introduce instead of $\vartheta(x)$
a new quantum field $\vartheta(x;\eta)$ defined by
\begin{eqnarray}\label{label3.17}
\hspace{-0.3in}\vartheta(x;\eta) &=&\frac{1}{2\pi}
\int^{\infty}_{-\infty}\frac{dk^1}{2k^0}\,\Big(a(k^1)e^{\textstyle
-i\,k\cdot x} + a^{\dagger}(k^1)e^{\textstyle i\,k\cdot
x}\Big)\nonumber\\ &+&\frac{1}{2\pi}
\int^{\infty}_{-\infty}dk^1\,\eta(k^1 - \pi M)\sqrt{\frac{\cos((k^1 -
\pi M)\lambda_{\rm M})}{2k^0|k^1 - \pi M|}}\,[a(k^1) +
a^{\dagger}(k^1)].
\end{eqnarray}
Since the $\eta(k^1)$ is tangible only in the region of momenta $k^1$
comeasurable with zero, the shift of the argument $k^1 \to k^1 - \pi
M$ leads to a concentration of the integrand around the momenta
$k^1\simeq \pi M$, where $\cos((k^1 -\pi M)\lambda_{\rm M})$ is
positive. The contribution of the constant term in the definition of
the $\vartheta$--field demonstrates the fact that the
$\vartheta$--field can be excited above the background by quanta with
momenta of order of the scale $M$.

Averaging over the $\eta$--field fluctuations we obtain the quantum
field $\vartheta(x)$ defined by (\ref{label1.7})
\begin{eqnarray}\label{label3.18}
\langle \vartheta(x;\eta)\rangle = \int{\cal
D}\eta\,\vartheta(x;\eta)\,\exp\Big\{-\frac{1}{2\pi}
\int^{\infty}_{-\infty}\frac{dk^1}{2k^0}\,\eta^2(k^1)\Big\}
= \vartheta(x).
\end{eqnarray}
Due the constraint (\ref{label2.9}) the additional contribution in
(\ref{label3.17}), independing on space--time coordinates and
containing the random $\eta(k^1)$--field, does not change the
generating functional of Green functions (\ref{label2.20}). However,
this contribution turns out to be important for the derivation of the
term $\cos(k^1\lambda_{\rm M})$ in the regularized two--point Wightman
functions (\ref{label2.25}), represented in terms of vacuum
expectation values of the products of the $\vartheta(x;\eta)$--fields
and averaged over the random $\eta$--field fluctuations. The
regularized two--point Wightman functions $D^{(\pm)}(x;M)$ we define
as follows
\begin{eqnarray}\label{label3.19}
\hspace{-0.3in}&&D^{(+)}(x;M)= \nonumber\\
\hspace{-0.3in}&&=\int{\cal D}\eta\,\langle
0|\vartheta(x;\eta)\vartheta(0;\eta)|0\rangle\,
\exp\Big\{-\frac{1}{2\pi} \int^{\infty}_{-\infty}\frac{dk^1}{2k^0}\,
\eta^2(k^1)\Big\}=\frac{1}{2\pi}\int^{\infty}_{-\infty}
\frac{dk^1}{2k^0}\,
e^{\textstyle -\,i\,k\cdot x}\nonumber\\
\hspace{-0.3in}&&+
\frac{1}{2\pi}\int^{\infty}_{-\infty}dk^1\sqrt{\frac{\cos((k^1 - \pi
M)\lambda_{\rm M})}{2k^0|k^1 - \pi M|}}
\frac{1}{2\pi}\int^{\infty}_{-\infty}dq^1\sqrt{\frac{\cos((q^1 - \pi
M)\lambda_{\rm M})}{2q^0|q^1 - \pi M|}}\nonumber\\
\hspace{-0.3in}&&\times\int{\cal D}\eta\,\eta(k^1 - \pi M)\, \eta(q^1
- \pi M )\,\exp\Big\{-\frac{1}{2\pi}
\int^{\infty}_{-\infty}\frac{dp^1}{2p^0}\,
\eta^2(p^1)\Big\},\nonumber\\ \hspace{-0.3in}&&D^{(-)}(x;M)= \nonumber\\
\hspace{-0.3in}&&=\int{\cal D}\eta\,\langle
0|\vartheta(0;\eta)\vartheta(x;\eta)|0\rangle\,
\exp\Big\{-\frac{1}{2\pi} \int^{\infty}_{-\infty}\frac{dk^1}{2k^0}\,
\eta^2(k^1)\Big\}= \frac{1}{2\pi}\int^{\infty}_{-\infty}
\frac{dk^1}{2k^0}\,e^{\textstyle +\,i\,k\cdot x}\nonumber\\
\hspace{-0.3in}&& +
\frac{1}{2\pi}\int^{\infty}_{-\infty}dk^1\sqrt{\frac{\cos((k^1 - \pi
M)\lambda_{\rm M})}{2k^0|k^1 - \pi M|}}\,\frac{1}{2\pi}
\int^{\infty}_{-\infty}dq^1\sqrt{\frac{\cos((q^1 - \pi M)\lambda_{\rm
M})}{2q^0|q^1 - \pi M|}}\nonumber\\
\hspace{-0.3in}&&\times\int{\cal D}\eta\,\eta(k^1 - \pi M)\, \eta(q^1
- \pi M)\,\exp\Big\{-\frac{1}{2\pi}
\int^{\infty}_{-\infty}\frac{dp^1}{2p^0}\, \eta^2(p^1)\Big\}.
\end{eqnarray}
For the integration over the $\eta$--field fluctuations we use the
auxiliary integrals
\begin{eqnarray}\label{label3.20}
&&\int{\cal D}\eta\,\exp\Big\{-\frac{1}{2\pi}
\int^{\infty}_{-\infty}\frac{dp^1}{2p^0}\, \eta^2(p^1) +
\frac{1}{\pi}\int^{\infty}_{-\infty}
\frac{dp^1}{2p^0}\,f(p^1)\,\eta(p^1)\Big\} = \nonumber\\
&&=\exp\Big\{\frac{1}{2\pi}\int^{\infty}_{-\infty}
\frac{dp^1}{2p^0}\,f^2(p^1)\Big\}
\end{eqnarray}
The integration of the quadratic terms in the $\eta$--field in
(\ref{label3.19}) we perform with the help of a formula which can be
derived from Eq.(\ref{label3.20})
\begin{eqnarray}\label{label3.21}
\hspace{-0.5in}&&\int{\cal D}\eta\,
\exp\Big\{-\frac{1}{2\pi}\int^{\infty}_{-\infty}\frac{dp^1}{2p^0}\,
\eta^2(p^1)\Big\}\,\eta(k^1 - \pi M)\, \eta(q^1 - \pi M)
= \frac{1}{4}\,\frac{\delta}{\delta f(k^1 - \pi
M)}\nonumber\\
\hspace{-0.5in}&&\times\,\frac{\delta}{\delta f(q^1 - \pi
M)}\exp\Big\{\frac{1}{2\pi}\int^{\infty}_{-\infty}
\frac{dp^1}{2p^0}\,f^2(p^1)\Big\}\Bigg|_{\textstyle f=0}=
(2\pi)\,|k^1 - \pi M|\,\delta(k^1 - q^1),
\end{eqnarray}
where we have used the relations
\begin{eqnarray}\label{label3.22}
\frac{\delta f(q^1}{\delta f(k^1)} &=& (2\pi)\,2k^0\,\delta(k^1 -
q^1),\nonumber\\ \frac{\delta f(q^1 - \pi M)}{\delta f(k^1- \pi M)}
&=& (2\pi)\,2|k^1 - \pi M|\,\delta(k^1 - q^1).
\end{eqnarray}
The regularized two--point Wightman functions read
\begin{eqnarray}\label{label3.23}
D^{(+)}(x;M) &=&
\frac{1}{2\pi}\int^{\infty}_{-\infty}\frac{dk^1}{2k^0}\,
\Big(e^{\textstyle -\,i\,k\cdot x} - \cos(k^1\lambda_{\rm
M})\Big),\nonumber\\ D^{(-)}(x;M)
&=&\frac{1}{2\pi}\int^{\infty}_{-\infty}\frac{dk^1}{2k^0}\,
\Big(e^{\textstyle +\,i\,k\cdot x} - \cos(k^1\lambda_{\rm M})\Big).
\end{eqnarray}
This testifies the removal of infrared divergences in the quantum
field theory of the free massless (pseudo)scalar field in
1+1--dimensional space--time.

\section{Conclusion}
\setcounter{equation}{0}

\hspace{0.2in} We have shown that the quantum field theory of a free
massless (pseudo)scalar field in 1+1--dimensional space--time does not
really suffer from an {\it infrared} problem. The generating
functional of Green functions $Z[J]$ given by (\ref{label2.20})
allowing to calculate any correlation function in the quantum field
theory of the free massless (pseudo)scalar field $\vartheta(x)$ does
not depend on the infrared cut--off. This occurs due to a simple
property of the external source of the free massless pseudoscalar
field to have a vanishing integral over the 1+1--dimensional volume
(\ref{label2.9}). The physical meaning of the constraint
(\ref{label2.9}) can be easily clarified in the momentum
representation, where it corresponds to a removal of a zero--mode
configuration of the free massless (pseudo)scalar field
$\vartheta(x)$.  According to a mechanical analogy of the free
massless (pseudo)scalar field $\vartheta(x)$ as a chain of $N$ linear
self--coupled harmonic oscillators, a zero--mode configuration of this
system describes a motion of a center of mass which does not affect
the evolution of the system and can be removed from the system without
a loss of information about the response of the system to external
perturbations induced by external sources obeying the constraint
(\ref{label2.9}).

An analogous treatment of a zero--mode configuration of massless
 self--coupled scalar fields, described by the $\sigma$--models with
 internal $O(N)$ symmetry and defined in one and two dimensions, has
 been suggested by Hasenfratz [21]. Hasenfratz has shown that the
 exclusion of the zero--mode configuration has allowed to construct a
 consistent perturbation theory with correct Feynman rules.

By virtue of the constraint (\ref{label2.9}) the infrared cut--off
$\mu$ in the generating functional of Green functions (\ref{label2.4})
can be replaced by an arbitrary finite scale $M$. The dependence of
$Z[J]$ on a finite scale $M$ allows to regularize the causal
two--point Green function of the $\vartheta$--field as well as
Wightman functions in the infrared region.

For the regularization of Wightman functions at the level of a
redefinition of the $\vartheta$--field we have used the technique of
coherent states. We have assumed that the low--frequency quanta of a
free massless (pseudo)scalar field are randomized and the wave
function of the system of low--frequency quanta can be described as a
coherent state. The coherent state of low--frequency quanta depends on
a randomized field $\eta(k^1)$ having tangible values only in the
close vicinity of zero momenta.  Treating this random field as a
hidden parameter of an ensemble of low--frequency quanta of the free
massless (pseudo)scalar field $\vartheta(x)$ we have assumed that the
regularized quantities should be obtained by averaging over the
$\eta$--field fluctuations.

Following this prescription we have introduced a new field operator
$\vartheta(x;\eta)$ (\ref{label3.17}) containing translationary
invariant quantum contribution to the standard $\vartheta$--field
expanded into plane waves (\ref{label1.7}). This quantum contribution
is proportional to the random $\eta$--field and vanishes when averaged
over the $\eta$--field fluctuations. Defining Wightman functions in
terms of the new quantum field operators $\vartheta(x;\eta)$ and
averaging over the $\eta$--field fluctuations we have arrived at
infrared convergent functions $D^{(\pm)}(x;M)$ depending on a finite
scale $M$. This solves the problem of a consistent, free of infrared
divergences, definition of a free massless (pseudo)scalar field in
1+1--dimensional space--time formulated by Klaiber [1] and Coleman
[4].

In order to discuss the relation of the free massless (pseudo)scalar
field theory free of infrared divergences to the MWH theorem we
suggest to calculate the vacuum expectation values of the operators
$e^{\textstyle +\,i\,\beta\,\vartheta(x)}$ and $e^{\textstyle
+\,i\,\beta\,[\vartheta(x) - \vartheta(y)]}$. The calculation runs in
the way [5,12--16]
\begin{eqnarray}\label{label4.1}
\hspace{-0.3in}\Big\langle 0\Big|e^{\textstyle
i\,\beta\,\vartheta(x)}\Big|0\Big\rangle =\Big\langle
0\Big|e^{\textstyle i\,\beta\,\vartheta(0)}\Big|0\Big\rangle =
\exp\Big\{\beta\,\frac{\delta}{\delta J(0)}\Big\}Z[J]\Big|_{\textstyle
J=0} = e^{\textstyle \frac{1}{2}\,\beta^2\,i\Delta(0;M)}.
\end{eqnarray}
The vacuum expectation value of the operator $e^{\textstyle
+\,i\,\beta\,[\vartheta(x) - \vartheta(y)]}$
can be calculated in analogous way and is equal to
\begin{eqnarray}\label{label4.2}
\hspace{-0.5in}\Big\langle 0\Big|{\rm T}\Big(e^{\textstyle
+\,i\,\beta\,[\vartheta(x) - \vartheta(y)]}\Big)\Big|0\Big\rangle =
\Bigg[e^{\textstyle \frac{1}{2}\,\beta^2\,i\Delta(0;M)}\Bigg]^2
e^{\textstyle -\,\beta^2\,i\Delta(x-y;M)}.
\end{eqnarray}
In the infrared singular quantum field theory of the free massless
(pseudo)scalar field $\vartheta(x)$ the causal Green function
$i\Delta(0;\mu)$ amounts to [5,12--16]
\begin{eqnarray}\label{label4.3}
i\Delta(0;\mu)=-\frac{1}{4\pi}\,{\ell
n}\Bigg(\frac{\Lambda^2}{\mu^2}\Bigg),
\end{eqnarray}
where $\Lambda$ is the ultra--violet cut--off.

Due to this the correlation function (\ref{label4.1}) in the infrared
singular quantum field theory of the free massless (pseudo)scalar
field $\vartheta(x)$ should read [5,12--16]
\begin{eqnarray}\label{label4.4}
\Big\langle 0\Big|e^{\textstyle
i\,\beta\,\vartheta(x)}\Big|0\Big\rangle = \Big\langle
0\Big|e^{\textstyle i\,\beta\,\vartheta(0)}\Big|0\Big\rangle
=\lim_{\textstyle \mu \to 0}
\Bigg(\frac{\mu^2}{\Lambda^2}\Bigg)^{\textstyle \beta^2/8\pi} = 0.
\end{eqnarray}
The r.h.s. of (\ref{label4.4}) vanishes for any value of the
ultra--violet cut--off $\Lambda$. In turn, the vacuum expectation
value of the operator $e^{\textstyle +\,i\,\beta\,[\vartheta(x) -
\vartheta(y)]}$ calculated in the infrared singular quantum field
theory of the free massless (pseudo)scalar field $\vartheta(x)$
contains the difference of the Green functions $\Delta(0;\mu) -
\Delta(x-y;\mu)$ for which the infrared cut--off $\mu$ cancels itself
and the correlation function (\ref{label4.2}) turns out to be
independent of $\mu$ and is, therefore, finite in the limit $\mu \to
0$. This property has been used in the literature [7,13--15] to draw a
similarity between the MWH theorem and the infrared singular quantum
field theory of the free massless (pseudo)scalar field $\vartheta(x)$.

In the infrared non--singular quantum field theory of the free
massless (pseudo)scalar field $\vartheta(x)$ the factor $e^{\textstyle
\frac{1}{2}\,\beta^2\,i\Delta(0;M)}$ is equal to
\begin{eqnarray}\label{label4.5}
e^{\textstyle \frac{1}{2}\,\beta^2\,i\Delta(0;M)} = \lim_{\textstyle
\epsilon \to 0}(-M^2\epsilon^2 + i\,0)^{\textstyle \beta^2/8\pi} = 0.
\end{eqnarray}
Therefore, the vacuum expectation values (\ref{label4.1}) and
(\ref{label4.2}) vanish simultaneously due to the short--distance
behaviour, or differently the ultra--violet divergences. In turn,
according to Coleman's analysis of the sine--Gordon model [11]
ultra--violet divergences can be removed by renormalization. Following
Coleman [11] and removing the common factor (\ref{label4.5}) we obtain
the correlation functions
\begin{eqnarray}\label{label4.6}
\Big\langle 0\Big|e^{\textstyle
i\,\beta\,\vartheta(0)}\Big|0\Big\rangle &=& 1,\nonumber\\ \Big\langle
0\Big|{\rm T}\Big(e^{\textstyle +\,i\,\beta\,[\vartheta(x) -
\vartheta(y)]}\Big)\Big|0\Big\rangle &=&[-M^2(x-y)^2 +
i\,0]^{\textstyle - \beta^2/4\pi}.
\end{eqnarray}
The most general correlation function of the time--ordered product of
the operators $e^{\textstyle +\,i\,\beta\,\vartheta(x_j)}\,(j =
1,\ldots,n)$ and $e^{\textstyle -\,i\,\beta\,\vartheta(y_k)}\,(k =
1,\ldots,p)$ given by (\ref{label2.3}) is equal to
\begin{eqnarray}\label{label4.7}
\hspace{-0.7in}&&G(x_1,\ldots,x_n;y_1,\ldots,y_p)= \Big\langle
0\Big|{\rm T}\Big(\prod^n_{j=1}e^{\textstyle +
i\beta\vartheta(x_j)}\prod^p_{k=1}e^{\textstyle
-i\beta\vartheta(y_k)}\Big)\Big|0\Big\rangle=\nonumber\\
\hspace{-0.7in}&&=\exp\Big\{-i\beta\sum^n_{j = 1}\frac{\delta}{\delta
J(x_j)} + i\beta\sum^p_{k = 1}\frac{\delta}{\delta
J(y_k)}\Big\}Z[J]\Big|_{\textstyle J = 0} = \Bigg[e^{\textstyle
\frac{1}{2}\,\beta^2\,i\Delta(0;M)}\Bigg]^{n+p}\nonumber\\
\hspace{-0.7in}&&\times\,\exp\Big\{\beta^2\sum^{n}_{j < k}i\Delta(x_j
- x_k; M) + \beta^2\sum^{p}_{j < k}i\Delta(y_j - y_k; M) -
\beta^2\sum^{n}_{j = 1}\sum^{p}_{k = 1}i\Delta(x_j - y_k; M)\Big\}.
\end{eqnarray}
Removing the ultra--violet divergences by renormalization [11] we
arrive at the expression
\begin{eqnarray}\label{label4.8}
\hspace{-0.3in}&&G(x_1,\ldots,x_n;y_1,\ldots,y_p) = \Big\langle
0\Big|{\rm T}\Big(\prod^{n}_{j=1}e^{\textstyle
+\,i\,\beta\,\vartheta(x_j)} \prod^{p}_{k=1}e^{\textstyle
-\,i\,\beta\,\vartheta(y_k)}\Big)\Big|0\Big\rangle =\nonumber\\
\hspace{-0.3in}&&=\exp\Big\{\beta^2\sum^{n}_{j < k}i\Delta(x_j - x_k;
M) + \beta^2\sum^{p}_{j < k}i\Delta(y_j - y_k; M) - \beta^2\sum^{n}_{j
= 1}\sum^{p}_{k = 1}i\Delta(x_j - y_k; M)\Big\}=\nonumber\\
\hspace{-0.3in}&&= \frac{\displaystyle \prod^{n}_{j < k}
[-M^2(x_j-x_k)^2]^{\textstyle \beta^2/4\pi}\prod^{p}_{j < k}
[-M^2(y_j-y_k)^2]^{\textstyle \beta^2/4\pi}}{\displaystyle
\prod^{n}_{j = 1} \prod^{p}_{k = 1}[-M^2(x_j-y_k)^2]^{\textstyle
\beta^2/4\pi}}.
\end{eqnarray}
The most natural way of a simultaneous removal of the ultra--violet
divergences is the use of dimensional or analytical regularization
allowing to set $\Delta(0; M) = 0$ [5]. 

We would like to accentuate that since due to the constraint
(\ref{label2.9}) the generating functional of Green functions $Z[J]$
given by (\ref{label2.14}) (or (\ref{label2.20})) is invariant under
the scale transformation $M \to M\,'$ , the correlation functions
(\ref{label4.7}) are invariant under the scale transformation $M \to
M\,'$ too. Of course, after the renormalization of ultra--violet
divergences carried out at the fixed normalization scale $M$, the
correlation functions (\ref{label4.8}) depend on $M$. However, this is
normal for any massless quantum field theory, since the correlation
functions (\ref{label4.8}) cannot be measured directly. The measurable
quantities are the elements of the $S$--matrix. In a free massless
(pseudo)scalar field theory the $S$--matrix is trivial and equal to
unity $S = 1$. The former testifies the independence of the
$S$--matrix on the normalization scale $M$.

The unit value of the vacuum expectation value of the operator
$e^{\textstyle +\,i\,\beta\,\vartheta(0)}$ in (\ref{label4.6}) can be
explained as follows. Due to the regularization of the field theory in
the infrared region the $\vartheta$--field is not anymore randomized.
It does not acquire large {\it classical} values proportional to
$2\pi$, as has been pointed out in Refs.[7,15], but varies smoothly
around zero, compatible with vacuum fluctuations $\langle
0|\vartheta(0)|0\rangle = 0$.

The time--ordered correlation function of the operator $e^{\textstyle
+\,i\,\beta\,[\vartheta(x) - \vartheta(y)]}$ in (\ref{label4.6})
agrees completely with the results obtained by different authors
[12--15] and the results we got in Ref.[5] for the solution of the
massless and massive Thirring model with fermion fields quantized in
the chirally broken phase. Recall, that the chirally broken phase of
the massless Thirring model possesses the ground state [5] coinciding
fully with the ground state of the superconducting phase in the
Bardeen--Cooper--Schrieffer (BCS) theory of superconductivity [24].

Now let us show that our approach to a quantum field theory of a free
massless (pseudo)scalar field $\vartheta(x)$ does not contradict to
the MWH theorem. For this aim we suggest to turn to the proof of the
MWH theorem expounded in Ref.[13] (see Appendix 4.A, p.219). We have
to show that the inequality (A.7) of Ref.[13], transcribed in our
notations in continuous Euclidean space for a free massless
(pseudo)scalar field theory formulated above, is fulfilled.

In our notations Eq.(A.7) of Ref.[13] can be transcribed in continuous
Euclidean space as follows\,\footnote{This can be obtained by
multiplying the inequality above Eq.(A.7) of Ref.[13] by a factor
$e^{\textstyle i\vec{k}\cdot \vec{\rho}}$ and integrating over
$\vec{k}$ at the limit $H \to 0$, where $H$ is an external magnetic
field.}
\begin{eqnarray}\label{label4.9}
\beta^2[\langle 0|\cos\beta\vartheta(0)|0\rangle]^2
(-i)\,\Delta(\vec{\rho}\,; M) \le \langle
0|\sin\beta\vartheta(0)\sin\beta\vartheta(\vec{\rho}\,)|0\rangle,
\end{eqnarray}
where $\vec{\rho} = (\rho_1,\rho_2)$ is an infinitesimal
2--dimensional Euclidean vector, $\Delta(\vec{\rho}\,; M)$ is a Green
function given by
\begin{eqnarray}\label{label4.10}
- i\,\Delta(\vec{\rho}\,; M) = \frac{1}{2\pi}\,{\ell
n}\Bigg(\frac{1}{M\rho}\Bigg).
\end{eqnarray}
Here $\rho = \sqrt{\rho^2_1 + \rho^2_2}$ and $M$ is a finite scale.

The inequality (\ref{label4.9}) is self--consistent. This can be
verified in the limit $\beta \to 0$. In fact, in the limit $\beta \to
0$ the inequality (\ref{label4.9}) becomes the equality corresponding
to the definition of the Wightman function of a free massless
(pseudo)scalar field $\vartheta(\vec{\rho}\,)$ in a 2--dimensional
Euclidean space.

In order to show this we suggest to divide both sides of
(\ref{label4.9}) by $\beta^2$. Then, taking the limit $\beta \to 0$ we
arrive at the relation
\begin{eqnarray}\label{label4.11}
- i\,\Delta(\vec{\rho}\,; M) \le \langle
  0|\vartheta(0)\vartheta(\vec{\rho}\,)|0\rangle.
\end{eqnarray}
Since by definition $\langle
0|\vartheta(0)\vartheta(\vec{\rho}\,)|0\rangle = - (1/2\pi)\,{\ell
n}(M\rho)$ is the Wightman function $D(\vec{\rho}\,; M)$ of a free
massless (pseudo)scalar field $\vartheta(\vec{\rho}\,)$ in
2--dimensional Euclidean space (\ref{label2.19}), we are able to set
$D(\vec{\rho}\,; M) = - i\,\Delta(\vec{\rho}\,; M) = - (1/2\pi)\,{\ell
n}(M\rho)$. Therefore, the relation (\ref{label4.11}) should be
rewritten in the form of the equality
\begin{eqnarray}\label{label4.12}
D(\vec{\rho}\,; M) = - i\,\Delta(\vec{\rho}\,; M) = \langle
  0|\vartheta(0)\vartheta(\vec{\rho}\,)|0\rangle.
\end{eqnarray}
This confirms the self--consistency of the inequality
(\ref{label4.9}).

According to Itzykson and Drouffe [13] the vacuum expectation value
$\langle 0|\cos\beta\vartheta(0)|0\rangle$ should be identified with
the spontaneous {\it magnetization} ${\cal M}$
\begin{eqnarray}\label{label4.13}
{\cal M} = \langle 0|\cos\beta\vartheta(0)|0\rangle.
\end{eqnarray}
The vacuum expectation value of the r.h.s. of (\ref{label4.9}) is
equal to
\begin{eqnarray}\label{label4.14}
\langle
0|\sin\beta\vartheta(0)\sin\beta\vartheta(\vec{\rho}\,)|0\rangle =
\frac{1}{2}\,e^{\textstyle \beta^2\,i\Delta(0; M)}\Bigg[e^{\textstyle
- \beta^2\,i\Delta(\vec{\rho}\,; M)} - e^{\textstyle +
\beta^2\,i\Delta(\vec{\rho}\,; M)}\Bigg].
\end{eqnarray}
Using (\ref{label4.1}) for the calculation of the spontaneous {\it
magnetization } ${\cal M}$ and Eqs.(\ref{label4.13}) and
(\ref{label4.14}) we recast the inequality (\ref{label4.9}) into the
form
\begin{eqnarray}\label{label4.15}
e^{\textstyle \beta^2\,i\Delta(0;M)}\,{\ell
n}\Bigg(\frac{1}{M\rho}\Bigg)\le \frac{\pi}{\beta^2}\,e^{\textstyle
\beta^2\,i\Delta(0;M)}\Big[(M\rho)^{\textstyle -\beta^2/2\pi} -
(M\rho)^{\textstyle \beta^2/2\pi}\Big].
\end{eqnarray}
Renormalizing $\langle 0|\cos\beta\vartheta(0)|0\rangle$ and $\langle
0|\sin\beta\vartheta(0)\sin\beta\vartheta(\vec{\rho}\,)|0\rangle$,
i.e. removing the constant divergent factors $e^{\textstyle
\beta^2\,i\Delta(0;M)}$ in both sides of (\ref{label4.15}) and getting
${\cal M} = 1$, we obtain
\begin{eqnarray}\label{label4.16}
{\ell n}\Bigg(\frac{1}{M\rho}\Bigg)\le
\frac{\pi}{\beta^2}\,\Big[(M\rho)^{\textstyle -\beta^2/2\pi} -
(M\rho)^{\textstyle \beta^2/2\pi}\Big].
\end{eqnarray}
This inequality is always fulfilled for $\rho \to 0$ as required by
the derivation of the MWH theorem according to Ref.[13]. Thus, the
renormalized spontaneous {\it magnetization} ${\cal M} = 1$, obtained
in our approach as a vacuum expectation value of the operator
$\cos\beta\vartheta(0)$, agrees with the inequality (\ref{label4.9}),
which is a continuous analogy of the lattice inequality used for the
derivation of the MWN theorem in [13]. We notice that the renormalized
spontaneous {\it magnetization} ${\cal M} = 1$, being the observable
quantity, does not depend on the normalization scale $M$.

It is seen that in the limit $\beta \to 0$ the relation
(\ref{label4.16}) becomes an equality. This testifies the validity of
our statement given by (\ref{label4.12}).

We would like to emphasize that our result, ${\cal M} = 1$, does not
contradict to the MWH theorem, since it goes beyond the scope of the
applicability of the MWH theorem [10]. As has been pointed out by the
authors [10], the vanishing of the {\it long--range order} can be
inferred only for non--zero temperature $T\neq 0$ [10] and none
conclusion about its value can be derived for $T=0$ [10]. Since
spontaneous {\it magnetization} and fermion {\it condensation} have
been found in the free massless (pseudo)scalar field theory, discussed
above, and the massless Thirring model in Ref.[5] at $T = 0$, these
results cannot be considered as counterexamples to the MWH theorem.

\section*{Acknowledgement}

\hspace{0.2in} This work was supported in part by Fonds zur
F\"orderung der Wissenschaftlichen Forschung P13997-TPH.

\end{document}